\documentclass[%
 reprint,
 amsmath,amssymb,
 aps,
]{revtex4-2}

\usepackage[pagewise]{lineno} % line numbering package
\usepackage{microtype} % fancy typesetting package
\usepackage{graphicx}% Include figure files
\usepackage{dcolumn}% Align table columns on decimal point
\usepackage{bm}% bold math
\usepackage{color}
\definecolor{Blue}{rgb}{0,0,1}
\definecolor{Red}{rgb}{0,0,0}
\definecolor{Green}{rgb}{0,1,0}

\begin{document}

%\preprint{APS/123-QED}

\title{Controllable tunnelling of single flux-quanta mediated by quantum phase-slip in disordered superconducting loops}% Force line breaks with \\

\author{J.A. Potter}
 \email{jamie.potter@npl.co.uk \\ Present address: National Physical Laboratory, Teddington, TW11 0LW, United Kingdom}
\author{J.C. Fenton}%
\author{P.A. Warburton}
\affiliation{London Centre for Nanotechnology, UCL, 17-19 Gordon Street, London WC1H 0AH, United Kingdom}%


%\affiliation{
%\affiliation{
 %Third institution, the second for Charlie Author
%\affiliation{%
 %%This line break forced with \textbackslash\textbackslash
%}%


\date{\today}% It is always \today, today,

\begin{abstract}
Quantum phase-slip (QPS) is the exact dual to the well-known Josephson effect. Although there are numerous proposals for applications of QPS devices, experimental work to develop these remains in the relatively early stages. Significant barriers to exploiting QPS nanowires for useful technologies still exist, such as establishing robust nanowire fabrication methods that allow coupling to low-loss circuits, and demonstrating control over the QPS process with an experimenter-controlled external bias. Here we report experiments which show that both of these barriers have been overcome. We present measurements at 300 mK of NbN coplanar waveguide (CPW) resonators embedded with nanowires fabricated using a neon focused ion-beam. The internal quality factor exceeds $2\times10^{4}$ \textemdash{} significantly higher than previously reported in comparable experiments. The resonator frequency tunes periodically with an applied magnetic field, revealing tunnelling of the order parameter that always occurs at half-integer values of the applied flux. In contrast to previous studies of single QPS, the order-parameter tunnelling is shown to be adiabatic, demonstrating improved control over energy dissipation in nanowire QPS circuits. Our results highlight a promising pathway towards realising low-loss nanowire-based QPS devices.
%\begin{description}
%\end{description}
\end{abstract}

\maketitle

% \linenumbers

%\tableofcontents
%%%%%%%%%%%%%%%%%%%%%%%%%%%%%%%%%%%%%%%%%%%%%%%%%%%%%%%%%%%%%%%%%%%%%%%%%%%%%%%%%%%%%%%%%%%%%%%%%%%%%%%%%%%%%%%%%%%%%%%%%%%%%
\section{\label{sec:intro}Introduction}

Quantum circuits based on superconducting materials are currently the leading candidate for the implementation of a scalable quantum computer, already beginning to tackle relevant computation and simulation problems \citep{Arute_2020,QuantumShuttle_2020} and recently demonstrating `quantum advantage' \citep{QuantumSupremacy_2019}. The Josephson junction is near ubiquitous in these circuits, providing the necessary nonlinearity. A quantum phase-slip nanowire is the flux-charge dual to the Josephson junction \citep{Mooij_2006}, and in theory every Josephson junction-based circuit has a QPS dual. As well as being proposed as a qubit with favourable properties over traditional Josephson-based technology \citep{Harmans_2005,Stace_2019}, the QPS nanowire's dual property of a nonlinear quantum capacitance enables potential applications such as novel qubit-qubit couplers \citep{Kerman_2019}, parametric amplification for qubit readout, and a primary quantum current standard \citep{CurrentStandard_2019}. As yet, the huge potential of QPS nanowires remains to be fully exploited, and two key reasons for this stand out. Firstly, consistent and reliable fabrication of materials and nanowires with the requisite properties remains challenging; secondly, full control over individual QPS events in a nanowire has not yet been demonstrated.

The QPS phenomenon is most pronounced in quasi-one-dimensional superconducting nanowires, by which we mean that the cross-sectional dimensions of the nanowire are comparable to the coherence length $\xi$ \citep{Bezryadin_2008}. In these nanowires, quantum (or indeed thermal) fluctuations can lead to complete suppression of the superconducting order parameter over the cross-section of the wire. This in turn leads to a sudden change of $2\pi$ in the phase difference between the two ends of the wire, accompanied by the transfer of a quantised amount of magnetic flux equal to the magnetic flux quantum $\Phi_{0}=h/2e$ through the wire. This tunnelling of a flux quantum can be coherent \citep{Astafiev_2012,Peltonen_2013,Peltonen_2016} or incoherent \citep{Fenton_2018}, depending on the relative scales of the phase-slip energy $E_{\mathrm{S}}$ and the inductive energy $E_{\mathrm{L}}$ of the nanowire. When $E_{\mathrm{S}}/E_{\mathrm{L}}<1$, the magnetic flux quantum number is well-defined, and incoherent transfer of individual flux-quanta is observable. This is in direct analogy to small Josephson junctions with large charging energy, where tunnelling of single Cooper pairs can be observed \citep{Geerligs_1990}. It is the incoherent QPS regime that is the focus of this paper. Incoherent QPS occur probabilistically, with a frequency characterised by the phase-slip rate $\Gamma_{\mathrm{S}}\equiv E_{\mathrm{S}}/h$. However if one waits for a time $\tau$ that is much longer than $1/\Gamma_{\mathrm{S}}$, then flux-quantum tunnelling will occur with extremely high likelihood. A device displaying deterministic transfer of quantised amounts of magnetic flux may find useful applications in tasks such as energy-efficient classical digital logic processing \citep{Likharev_1991,Mukhanov_2011,Herr_2011,Kirichenko_2011,Volkmann_2012}. Development of such a device is a key enabling technology for control of superconducting quantum processors at technologically-useful scales.

Historically, phase slips were studied in externally connected, current-biased nanowires, where the collective effect of many phase slips manifests as a resistance below $T_{\mathrm{c}}$ \citep{Lukens_1970,Newbower_1972,VanRun_1987,Giordano_1988,Bezryadin_2000,Lau_2001}. However, in order to isolate a single phase-slip, it is necessary to incorporate the nanowire into a flux-biased superconducting loop. The flux-dependent energy states of a continuous superconducting loop are described by a set of parabolas (see Fig.~\ref{fig:DeviceFig}(a)), where each parabola corresponds to a unique value $N$ of the phase winding number, or equivalently the number of flux quanta in the loop. A single phase-slip corresponds to a transition between neighbouring parabolas, and if no lower-energy state is available at a particular external flux bias, then phase slips are forbidden. When sweeping the externally applied flux, tunnelling of a single flux-quantum becomes allowed at $\Phi_{\mathrm{app}} = (N+1/2)\Phi_{0}$ (known as the degeneracy point, highlighted in red in Fig.~\ref{fig:DeviceFig}(a)) \citep{Matveev_2002,Zhang_1997}. Tunnelling will occur when the system passes the degeneracy point if the rate of phase slips is much greater than the rate at which the flux is swept. However, if the flux is swept slowly with respect to $\Gamma_{\mathrm{S}}$, tunnelling will cause the system to enter a metastable state and it will then undergo irreversible relaxation to the ground state at some later time. A number of recent experiments \citep{Belkin_2011,Belkin_2015,Petkovic_2016,Petkovic_2019} have demonstrated relaxation via QPS from a higher-energy metastable state, but controlled single-flux-quantum tunnelling when passing through the degeneracy point has not previously been demonstrated.

In this paper, we demonstrate for the first time single-flux-quantum tunnelling occurring at the degeneracy point in a continuous superconducting loop. The flux quanta tunnel through NbN nanowires embedded in the loop and this is read out via coupling to a high-quality coplanar waveguide (CPW) resonator. An important innovation in our fabrication technique is that the nanowires were fabricated by neon focused-ion-beam (FIB) milling. The FIB process enables the repeatable fabrication of nanowires with width ${w\approx25\text{ nm}}$ ensuring a large phase-slip rate, while introducing minimal losses to the host resonator. FIB milling has previously been shown to be compatible with low-dissipation superconducting circuits \citep{Burnett_2016,Burnett_2017}; however, this is the first report of the use of FIB to fabricate a device in which quantum phase-slip behaviour has been measured. Our results show flux-periodic tuning of the resonant frequency $\nu_{0}$ while maintaining a high intrinsic quality factor $Q_{\mathrm{i}}$ at all values of applied flux. We show that a single quantum phase-slip always occurs when adjacent winding-number states become degenerate. This is ensured by a phase-slip rate \textemdash{} we estimate ${\Gamma_{\mathrm{S}} = 35\text{ MHz}}$ \textemdash{} which is large in comparison to $1/\tau_{\mathrm{E}}$, where $\tau_{\mathrm{E}}$ is the experimental timescale. However, the phase-slip rate is less than the inductive energy ($E_{\mathrm{L}}/h>1\text{ THz}$), and less than the thermal energy ($k_{\mathrm{B}}T/h \approx 6$ GHz), which excludes the possibility of avoided crossings associated with coherent QPS.

Our demonstration of the ability to control the tunnelling of single flux-quanta represents important progress towards applications of QPS devices in quantum information processing that have been proposed elsewhere. The low loss in our device suggests the potential for high tunnelling rates in QPS devices without a significant increase in $T_{1}$-type decoherence.  In addition to this, there is potential for a QPS digital logic processing device, based on the deterministic transfer of single flux-quanta \citep{Likharev_1991,Mukhanov_2011,Herr_2011,Kirichenko_2011}. Utilising quantum tunnelling of flux in such a device should enable significant reduction in the heat dissipation associated with each gate \citep{McDermott_2018}, a reduction that will be necessary for the scaling up of systems beyond the 1,000-qubit level.

%%%%%%%%%%%%%%%%%%%%%%%%%%%%%%%%%%%%%%%%%%%%%%%%%%%%%%%%%%%%%%%%%%%%%%%%%%%%%%%%%%%%%%%%%%%%%%%%%%%%%%%%%%%%%%%%%%%%%%%%%%%%%
\section{\label{sec:methods}Fabrication and Experimental Details}

\begin{figure}[t]
\begin{center}
\includegraphics[width=3.4in,keepaspectratio]{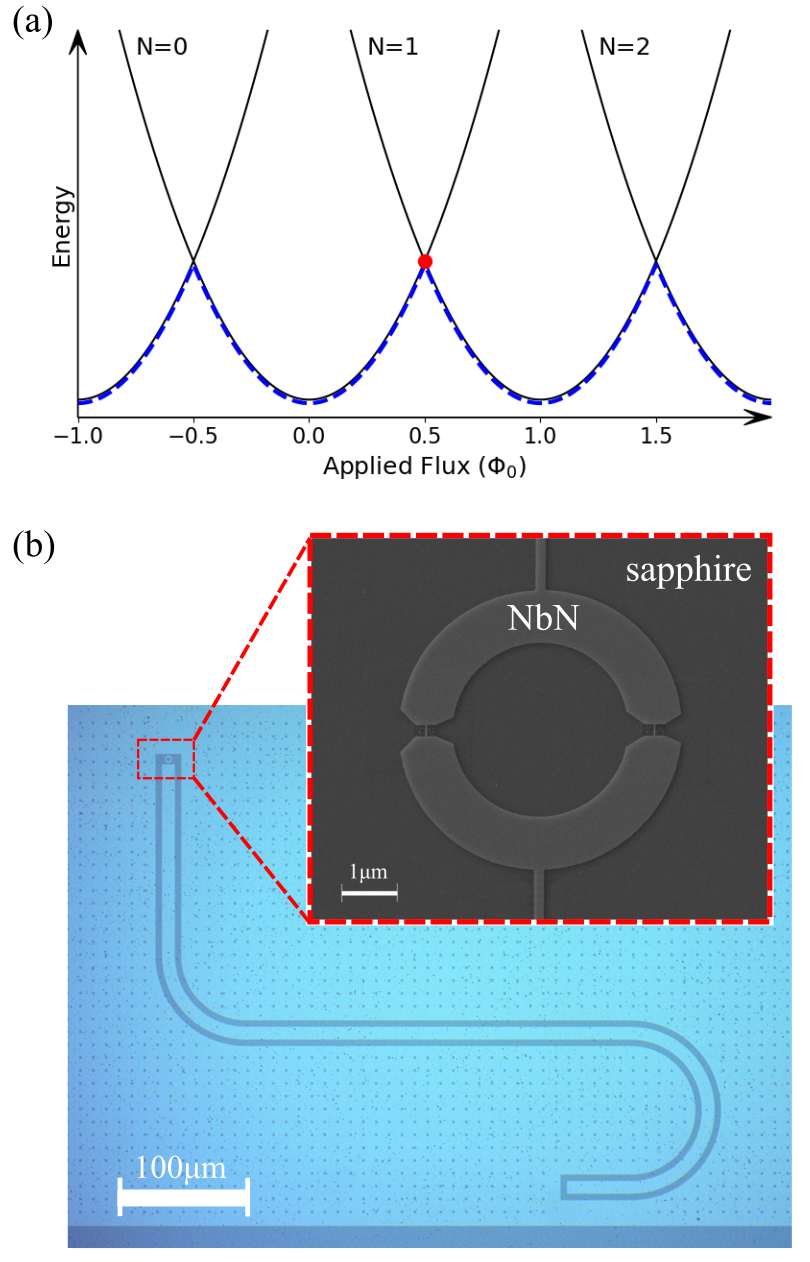}
\caption{\textbf{(a)} Flux-dependent energy spectrum of a continuous superconducting loop, with blue dashed line highlighting the ground state. A single flux-quantum may tunnel into the loop at the degeneracy point --- highlighted in red. $N$ is the winding number and $\Phi_{0}$ the flux quantum. \textbf{(b)} Optical (main image; blue contrast) and SEM (inset; grey contrast) images of the NbN nanowire-embedded loop located at the short-circuited termination of the CPW resonator. The nanowires in this device are 25-nm wide and 200-nm long. The lower and upper leads to the loop are connected to the CPW centre conductor and superconducting ground plane respectively.}
\label{fig:DeviceFig}
\end{center}
\end{figure}

We fabricated nanowire-embedded resonators from 10-nm-thick films of superconducting NbN. The NbN was deposited on $c$-axis oriented sapphire substrates by dc magnetron sputtering of a 99.99$\%$ pure niobium target in a 1:1 Ar:N$_{2}$ atmosphere at a pressure of $5 \times 10^{-3}$ mbar and a power of 150 W. The resulting film was measured to have critical temperature $T_{\mathrm{c}} = 8.55$ K and sheet resistance $R_{\mathrm{sq}} = 1.2$ k$\Omega$/sq.

\begin{figure*}[t]
\begin{center}
\includegraphics[width=6.8in,keepaspectratio]{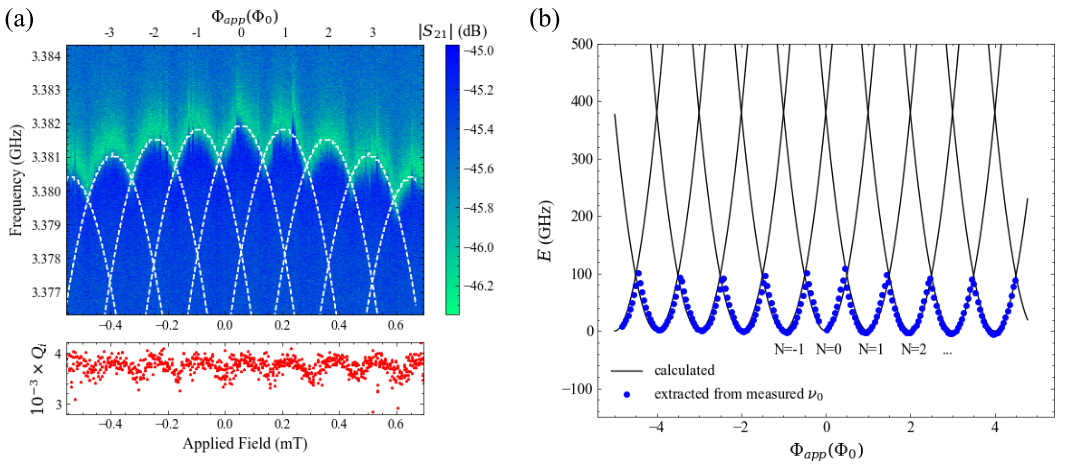}
\caption{\textbf{(a)} Upper panel: Single-tone spectroscopy of nanowire-embedded resonator measured at $T=305$ mK and $\langle n \rangle \approx 50$. $|S_{21}|$ is plotted as a function of frequency and applied magnetic field. The top axis shows the applied magnetic flux $\Phi_{\mathrm{app}}$ seen by the nanowire loop, which is inferred from the periodicity of the resonator tuning. Dashed white lines show resonant frequencies corresponding to the calculated energy states of the loop. Lower panel: Magnetic-field dependence of the measured intrinsic quality factor. \textbf{(b)} Calculated energy states of the loop and values extracted from the measured resonant frequency.}
\label{fig:ResultsFig}
\end{center}
\end{figure*}

Quarter-wavelength resonators were patterned by electron-beam lithography (EBL) into a 300-nm layer of PMMA resist. Multiple resonators on each chip are capacitively coupled to a common feedline and are patterned with narrow loops galvanically coupled at the short-circuited end. At this stage, the loops contain `precursor' nanowires designed with a width of 400 nm. The pattern was transferred into the NbN film by reactive ion etching (RIE) using a 2:5 volume ratio of CHF$_{3}$ and SF$_{6}$ at 100 W and 100 mbar.

Nanowires are then patterned into the loops using a neon focused-ion-beam, whereby a beam of Ne ions is accelerated from an atomically-defined tip onto the sample surface with spot size down to 2 nm and sufficient energy to sputter the metal film \citep{Nanopatterning_2005,CircuitEdit_2014}. The precursor wires were milled to a width of 25 nm (see Fig.~\ref{fig:DeviceFig}(c)) using a 15-keV Ne beam. A Ne dose of 0.5 nC/$\mu$m$^{-2}$ was sufficient to clear the 10-nm film.

The sample was wire-bonded to a copper printed-circuit-board (PCB) and enclosed within an ECCOSORB-lined brass sample box. This was cooled to a temperature of $T \sim 300$ mK using a $^{3}$He refrigerator. Measurement of the rf response of the device was made using a Vector Network Analyser (VNA) via an input line with ${60\text{ dB}}$ of attenuation to reduce thermal noise from room temperature. Signals in the output line were amplified with a high-electron-mobility-transistor (HEMT) amplifier. Global magnetic field was applied perpendicular to the plane of the loop using a superconducting solenoid and a precision current source. The lines supplying current to the coil were filtered at room temperature with an upper cut-off frequency of 9.2 kHz.

%%%%%%%%%%%%%%%%%%%%%%%%%%%%%%%%%%%%%%%%%%%%%%%%%%%%%%%%%%%%%%%%%%%%%%%%%%%%%%%%%%%%%%%%%%%%%%%%%%%%%%%%%%%%%%%%%%%%%%%%%%%%
\section{\label{sec:results}Results and Discussion}

\noindent\textbf{A. Flux Dependence of Resonant Frequency}

In this paper, we present results on a single NbN nanowire-embedded CPW resonator (see Supplemental Information for comparison of multiple devices). We measured the forward transmission ($S_{21}$) through the on-chip feedline, where the $\lambda/4$ resonators appear as a notch-type resonance. The upper panel of Fig.~\ref{fig:ResultsFig}(a) shows the main result of this work: under an applied magnetic field, the resonance tuning shows discontinuous changes of gradient at periodic values of the applied field. As we will demonstrate in the remainder of this paper, these discontinuities occur when two stable states with winding number differing by one become degenerate and are due to single-flux-quantum tunnelling mediated by quantum phase-slips in the nanowire \citep{Arutyunov_2008}.

The magnetic-field periodicity is 153 $\mu$T, which corresponds to a single flux-quantum in our loop assuming a flux-focusing factor $\mathcal{F} = 1.7$ \citep{FluxFocusing_2017}. The data in Fig.~\ref{fig:ResultsFig}(a) corresponds to a single direction of magnetic field sweep, but sweeps in the opposite direction were found to give the same result. We also observe a non-periodic, parabolic decrease of the resonant frequency as the magnitude of the applied field is increased. This is the expected \citep{Healey_2008,Zollitsch_2019} kinetic-inductance tuning of the NbN resonator, and can be parametrised by a phenomenological field-scale ${B_{\star} = 8\text{ mT}}$.

Figure~\ref{fig:ResultsFig}(a) also shows periodic variation of $Q_{\mathrm{i}}$ of the resonance as a function of applied field with the same field period as the resonant frequency. Quality factors were obtained by an analytical fit \citep{Probst_2015} to the equation
\begin{equation} \label{eq:ResFit}
    |S_{21}(\nu)| = ae^{i\alpha}e^{-i2\pi \nu\tau}\Bigg[1-\frac{\Big(\frac{Q_{\mathrm{l}}}{|Q_{\mathrm{c}}|}\Big)e^{i\phi}}{1+2iQ_{\mathrm{l}}(\nu/\nu_{0}-1)}\Bigg],    
\end{equation}
where $\nu$ is the probe frequency and $\nu_{0}$ is the resonance frequency. $Q_{\mathrm{c}}$ and $Q_{\mathrm{l}}$ are the coupling and loaded quality factors respectively, and obey the relationship $1/Q_{\mathrm{l}} = 1/Q_{\mathrm{i}} + 1/Q_{\mathrm{c}}$. $\phi$ accounts for the effect of impedance mismatches in the circuit, the scale factor $a$ represents the change in amplitude due to any attenuation/amplification in the measurement chain, $\alpha$ describes any initial phase offset of the signal, and $\tau$ accounts for frequency-dependent cable delay. We find that $Q_{\mathrm{i}}$ decreases approximately quadratically from $4\times10^{3}$ at $\delta\Phi \equiv (\Phi_{\mathrm{app}} - N\Phi_{0})/\Phi_{0} = 0$ to $3.4\times10^{3}$ at $\delta\Phi = 1/2$. We attribute this small change to non-equilibrium quasiparticles excited by the induced screening current providing an extra loss mechanism \citep{Levine_1965}. We observe no sharp decrease in $Q_{\mathrm{i}}$ at $\delta\Phi=1/2$, which suggests that the heat dissipated by the quantum phase-slip itself is not large enough to cause extra losses in the resonator. We note that the intrinsic quality factor exceeds any currently reported in the literature for QPS devices, and discuss this further in the Supplemental Information.

The periodic tuning of the resonance is well fitted by a model of an inductive superconducting loop remaining in its ground state (see Fig.~\ref{fig:ResultsFig}(b)), where the system is allowed to move between adjacent parabolas by transferring a single flux-quantum through the nanowire at ${\Phi_{\mathrm{app}} = (N+1/2)\Phi_{0}}$.

The loop is made up of a wide section and a narrow section (as shown in Fig.~\ref{fig:DeviceFig}(b)), and so can be modelled as two nonlinear kinetic inductances in series. The flux-dependent kinetic inductance of the loop $L_{\mathrm{k}}(\Phi)$ is therefore
\begin{equation} \label{eq:Lk}
    L_{\mathrm{k}}(I)=L_{\mathrm{k,1}}(0)\Bigg[1+\Bigg(\frac{I_{\mathrm{s}}}{I_{\star,1}}\Bigg)^2\Bigg]+L_{\mathrm{k,2}}(0)\Bigg[1+\Bigg(\frac{I_{\mathrm{s}}}{I_{\star,2}}\Bigg)^2\Bigg],
\end{equation}
where $I_{\star,1}$ and $I_{\star,2}$ are known to be of the order of the critical current in the wide and narrow section respectively \citep{Zmuidzinas_2012}. Since the screening current is $I_{\mathrm{s}}=\Phi/L_{\mathrm{k}}$, we can insert this into Eq.~\ref{eq:Lk} and solve for $L_{\mathrm{k}}(\Phi)$.

The input impedance $Z_{\mathrm{in}}$ of a $\lambda/4$ CPW resonator terminated by an inductive load, as a function of frequency and load impedance, is
\begin{equation} \label{eq:impedance}
    Z_{\mathrm{in}}(\nu,Z_{\mathrm{L}}) = Z_{0}\frac{Z_{\mathrm{L}} + iZ_{0}\tan{\big(\frac{2\pi\nu l}{c}\big)}}{Z_{0} + iZ_{L}\tan{\big(\frac{2\pi\nu l}{c}\big)}},
\end{equation}
where $Z_{0}$ is the characteristic impedance of the resonator, $Z_{L}(\Phi) = i2\pi\nu L_{\mathrm{k}}(\Phi)$ is the impedance of the inductive load, $c$ is the speed of light in the resonator, and $l$ is its length. At resonance, $\mathrm{Im}\{Z_{\mathrm{in}}\}\rightarrow\infty$, so given $L_{\mathrm{k}}(\Phi)$ one can numerically calculate $\nu_{0}(\Phi)$, or given $\nu_{0}(\Phi)$ one can numerically calculate $L_{\mathrm{k}}(\Phi)$. 

One can also calculate the flux-dependent free energy of the loop $E(\Phi)$ from $L_{\mathrm{k}}(\Phi)$, using the relation
\begin{equation} \label{eq:loopE}
    L_{\mathrm{k}}^{-1} = \frac{d^{2}E}{d\Phi^{2}}.
\end{equation}
To obtain the free energy, we simply numerically integrate the inverse of the inductance twice with respect to flux.

We calculated $L_{\mathrm{k}}(\Phi)$ for our device from Eq.~\ref{eq:Lk} using only independently determined parameters. A critical current density of $J_{\mathrm{c}} = 4.4\times10^{5}$ Acm$^{-2}$ was obtained from a dc measurement of a track etched into the NbN film, and a sheet kinetic inductance of ${L_{\mathrm{sq}} = 0.34\text{ nH/sq}}$ was inferred from the zero-field $\nu_{0}$ of the resonator. The geometry of the loop was measured by SEM and this was used to calculate $L_{\mathrm{k}}(0)$ and $I_{\star}$ (we set $I_{\star}=I_{\mathrm{c}}$). We then calculated the white dashed lines in Fig.~\ref{fig:ResultsFig}(a) using Eq.~\ref{eq:impedance}, and the solid black lines in Fig.~\ref{fig:ResultsFig}(b) from Eq.~\ref{eq:loopE}. The blue points in Fig.~\ref{fig:ResultsFig}(b) were extracted from the measured $\nu_{0}$ using Eqs.~\ref{eq:impedance} and \ref{eq:loopE}.

%%%%%%%%%%%%%%%%%%%%%%%%%%%%%%%%%%%%%%%%%%%%%%%%%%%%%%%%%%%%%%%%%%%
% Discussion
%%%%%%%%%%%%%%%%%%%%%%%%%%%%%%%%%%%%%%%%%%%%%%%%%%%%%%%%%%%%%%%%%%%
\begin{figure}[t]
\begin{center}
\includegraphics[width=3.4in,keepaspectratio]{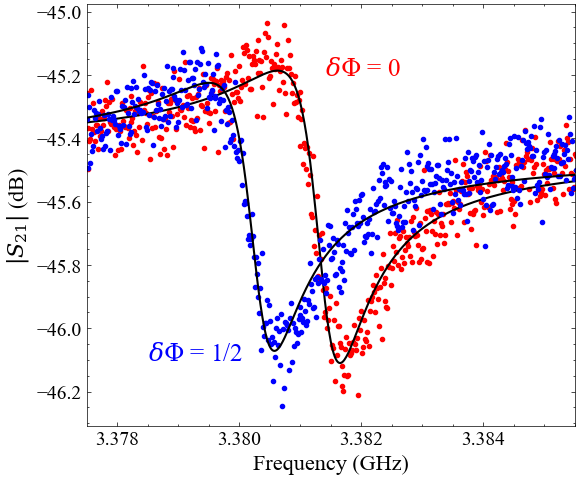}
\caption{$|S_{21}|$ measured at $\Phi_{\mathrm{app}}=0$ and $\Phi_{\mathrm{app}}=\Phi_{0}/2$ along with fit (black line) to a linear resonance model. This shows that at $\Phi_{\mathrm{app}}=\Phi_{0}/2$ the response is linear, so the current in the nanowire is well below the critical current.}
\label{fig:NonlinearityFig}
\end{center}
\end{figure}

\bigskip

\noindent\textbf{B. Mechanisms for Flux Quantum Transfer}

The periodic tuning of the resonator and the associated fit to the calculated energy states of the loop constitute strong evidence that the loop remains in its ground state for all values of $\Phi_{\mathrm{app}}$, and this is made possible by a single flux-quantum transferring into or out of the loop at $\delta\Phi = 1/2$. It is important to establish the mechanism by which the flux is able to enter the loop, so we now turn our attention to the transitions between flux states. Across multiple devices on multiple chips, we found an onset of flux-periodic tuning in devices containing nanowires with $w\lesssim35$ nm. This dependence of the behaviour on nanowire width suggests that the flux tunnelling occurs in the nanowires, and not in the wider part of the loop. We can now examine some possible physical processes that could occur in the nanowires and see how the data fits with them.

-\textit{Does the nanowire current exceed its critical current $I_{\mathrm{c}}^{\mathrm{nw}}$?} Niobium-nitride resonators commonly exhibit a nonlinear $S_{21}$ response when they are driven with a high microwave power \citep{Thomas_2020} as a result of the current-induced nonlinear kinetic inductance. As we see in Eq.~\ref{eq:Lk}, the kinetic inductance is quadratically dependent on $(I/I_{\mathrm{c}})^{2}$, and so the nonlinearity must be dominated by the part of the conductor with the lowest $I_{\mathrm{c}}$. This is confirmed by measurements of our circuits, where we find that nanowire-embedded resonators show a much higher degree of nonlinearity in their $S_{21}$ response than bare resonators (see Supplemental Information). The quadratic nature of the nonlinearity suggests that a strongly nonlinear response is a consequence of the magnitude of the rf current $I_{\mathrm{res}}$ in the resonator reaching a significant fraction of the nanowire critical current $I_{\mathrm{c}}^{\mathrm{nw}}$. Our NbN resonator readout method therefore gives us an indirect readout of whether the current in the nanowire is close to $I_{\mathrm{c}}^{\mathrm{nw}}$.

Figure~\ref{fig:NonlinearityFig} shows the $S_{21}$ response of the nanowire-embedded resonator at $\delta\Phi = 0$ and $\delta\Phi = 1/2$, both measured in the low-power limit with an estimated resonator photon population of $\langle n \rangle \approx 50$. In both cases, the response is linear and well fitted by Eq.~\ref{eq:ResFit}. We calculate, using the relation $I_{\mathrm{s}}=dE(\Phi)/d\Phi$ that the maximum induced screening current in the nanowires $I_{\mathrm{s}}(\delta\Phi=1/2)$ is 120 nA, an order of magnitude less than $I_{\mathrm{c}}^{\mathrm{nw}}$. Crucially, the lack of nonlinearity of the resonance at $\delta\Phi = 1/2$, along with the fact that $Q_{\mathrm{i}}$ remains a significant fraction of its zero-field value, means that the nanowires are not being driven close to their critical current by the applied flux. By a similar argument, we know that the nanowires are not being driven through $T_{\mathrm{c}}$ by a local heating process, as this would also result in nonlinearity of the resonance at $\delta\Phi = 1/2$ due to dissipation in the highly resistive normal metal.

-\textit{Is the nanowire a constriction Josephson-junction?} `Dayem-bridge' Josephson-junction SQUIDs are commonly embedded within CPW resonators \citep{SQUIDTuning_2008,FastTuning_2008} to provide a flux-tunable resonant frequency. However, when the SQUID loop has a large inductance, one observes hysteretic tuning, characterised by the parameter ${\beta_{L} = 2LI_{\mathrm{c}}/\Phi_{0}}$. When $\beta_{L}\gtrsim1$, the SQUID behaviour becomes hysteretic with applied flux and the resonator will exhibit discontinuous jumps in the resonant frequency, as observed in \citep{SQUIDHysteresis_2011,Kennedy_2019}. Our device does not undergo any discontinuous jumps, and the tuning over a single flux-quantum is symmetric, so $\beta_{L}<1$. Given this and the known loop inductance, we can set an upper bound on the critical current of any Josephson junction of ${I_{\mathrm{c}}^{\mathrm{JJ}}<100\text{ nA}}$. This bound is $10\times$ smaller than the expected transport critical current of our nanowires and also less than $I_{\mathrm{s}}(\delta\Phi_{0})$. Therefore, it is unrealistic to conclude that the flux-tuning we observe is a consequence of a dc SQUID formed of Dayem-bridge Josephson junctions. We also note that the closeness of the fit shown in Fig.~\ref{fig:ResultsFig} suggests there is no contribution to the flux-dependent inductance from a Josephson junction, which would add a $1/\cos{\Phi}$ term to Eq.~\ref{eq:Lk}.

\begin{figure}[t]
\begin{center}
\includegraphics[width=3.4in,keepaspectratio]{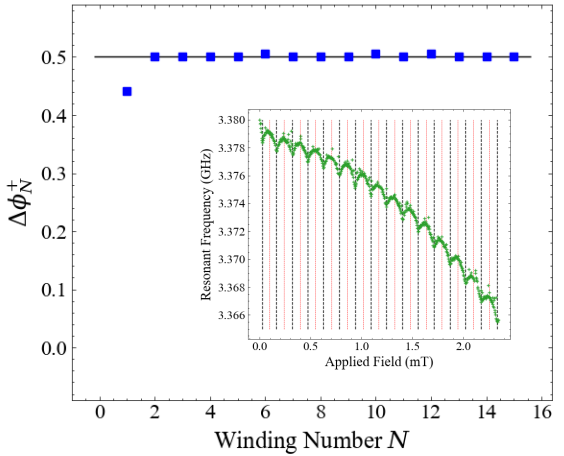}
\caption{Normalised tunnelling flux $\Delta\phi_{N}^{+}$ (defined in the main text), as a function of the winding number $N$. The solid line corresponds to half a flux quantum, $\Delta\phi_{N}^{+}=0.5$. Inset: Magnetic-field dependence of resonant frequency of the nanowire-embedded resonator up to an applied field of 2.4 mT. The black dotted lines mark transitions between winding number states ($\phi_{N}^{+}$), and the red dotted lines mark the energy minimum of each winding number state ($\phi_{{\mathrm{min}},N}$).}
\label{fig:HighBfig}
\end{center}
\end{figure}

-\textit{Is this a thermal or quantum process?} Figure~\ref{fig:HighBfig} shows that tunnelling always occurs at degeneracy. Following \citet{Petkovic_2016}, $\Delta\phi_{N}^{+}$ is defined as $\Delta\phi_{N}^{+}=\phi_{N}^{+}-\phi_{{\mathrm{min}},N}$, where $\phi_{N}^{+}$ is the normalised flux $\phi=\Phi/\Phi_{0}$ at which tunnelling from state $N$ to state $N+1$ occurs, and $\phi_{{\mathrm{min}},N}$ is the flux that minimises the loop free-energy for a particular winding number $N$. Our data shows the periodicity, defined in this way, to be half a flux quantum for all values of the winding number except $N=1$ (we attribute the latter exception to enhanced flux-focusing at low magnetic field). This is in contrast to \citep{Petkovic_2016}, where a characteristic dependence of $\Delta\phi_{N}^{+}$ on $N$ is shown to be a defining feature of thermally-activated phase-slips. By following the method of \citep{Petkovic_2016}, we calculate that $\Delta\phi_{N}^{+} \approx 300$ would be required in order for the energy barrier to phase slips in our nanowire to be tuned to $\lesssim k_{\mathrm{B}}T$. Correspondingly, we can estimate (see Supplemental Information) that ${\Gamma_{\mathrm{QPS}}=35\text{ MHz}}$ for our nanowire, and we calculate a temperature-dependent $\Gamma_{\mathrm{TAPS}}$ that is below ${1\text{ Hz}}$ for ${T<1.5\text{ K}}$. Therefore, at our measurement temperature, quantum tunnelling of flux is overwhelmingly more likely than a thermal transition. For comparison, the inverse experimental timescale is  ${1/\tau_{\mathrm{E}}\approx0.2\text{ Hz}}$ since, for each setpoint of the magnetic field, it takes the VNA approximately $5\text{ s}$ to collect $S_{21}$ data across the resonance. Because ${k_{B}T/h>\Gamma_{\mathrm{QPS}}\gg1/\tau_{\mathrm{E}}}$ \textemdash{} and to our knowledge this is the first reported study in this regime \textemdash~when our device bias is swept through the degeneracy point, a single quantum phase-slip always occurs before we are able to observe the system in a higher-energy metastable state.

%%%%%%%%%%%%%%%%%%%%%%%%%%%%%%%%%%%%%%%%%%%%%%%%%%%%%%%%%%%%%%%%%%%%%%%%%%%%%%%%%%%%%%%%%%%%%%%%%%%%%%%%%%%%%%%%%%%%%%%%%%%%%
\section{\label{sec:conclusion}Conclusions}

We have used a Ne FIB to fabricate NbN nanowires with widths down to 25 nm embedded within CPW resonators. We observe periodic modulation of resonant frequency and intrinsic quality factor, which is consistent with quantum tunnelling of individual flux quanta mediated by quantum phase-slip, occurring when states of different winding number become degenerate. This behaviour has been observed in resonators with intrinsic quality factor, $Q_{\mathrm{i}}$, up to $2.7\times10^{4}$ at 300 mK, which to our knowledge is the highest quality factor measured in quantum phase-slip experiments --- note that the losses here are significantly lower than suggested by comparable reports \citep{Astafiev_2012}. We estimate that the QPS rate is of the order 10\textendash100 MHz, which means that the tunnelling of a single flux-quantum is effectively deterministic on the timescale of microseconds. This result shows the suitability of the Ne FIB process for fabricating QPS devices. We also suggest that an incoherent QPS device with a high QPS rate such as ours could be promising for classical digital logic processing applications, where the quantum nature of the flux tunnelling implies a reduction in heat dissipation compared with current state-of-the-art devices, opening up a route to resolving an important roadblock to the upscaling of qubit control electronics.

%%%%%%%%%%%%%%%%%%%%%%%%%%%%%%%%%%%%%%%%%%%%%%%%%%%%%%%%%%%%%%%%%%%%%%%%%%%%%%%%%%%%%%%%%%%%%%%%%%%%%%%%%%%%%%%%%%%%%%%%%%%%%
\begin{acknowledgments}
The authors thank O.~W.~Kennedy for useful discussions. The authors gratefully acknowledge funding from the United Kingdom Engineering and Physical Sciences Research Council, Grant Nos.~EP/L015242/1, EP/J017329/1, and EP/T001062/1.
\end{acknowledgments}

\bibliography{references.bib}% Produces the bibliography via BibTeX.

\end{document}